\documentclass[preprint,12pt]{elsarticle}

\usepackage{amssymb}
\usepackage{amsmath}
\usepackage{mathtools}
\usepackage{algorithm,algpseudocode}
\journal{Acta Astronautica}

\begin{document}
	
\begin{frontmatter}
    
\title{Real-time rapid leakage estimation for deep space habitats using exponentially-weighted adaptively-refined search}

\author[inst1,inst2]{Mahindra Rautela}
\author[inst1]{Motahareh Mirfarah}
\author[inst1,inst3]{Christian Silva}
\author[inst1]{Shirley Dyke}
\author[inst1]{Amin Maghareh}
\author[inst2]{S.Gopalakrishnan}

\affiliation[inst1]{organization={Resilient Extraterrestrial Habitat Institute},
    addressline={Purdue University}, 
    city={West Lafayette},
    postcode={47906}, 
    state={Indiana},
    country={USA}}

\affiliation[inst2]{organization={Department of Aerospace Engineering},
    addressline={Indian Institute of Science}, 
    city={Bangalore},
    postcode={560012}, 
    state={Karnataka},
    country={India}}

\affiliation[inst3]{organization={Escuela Superior Politecnica del Litoral, ESPOL},
    addressline = {Km 30.5 Via Perimetral},
    city={Guayaquil},
    country={Ecuador}}

\begin{abstract}
The recent accelerated growth in space-related research and development activities makes the near-term need for long-term extraterrestrial habitats evident. Such habitats must operate under continuous disruptive conditions arising from extreme environments like meteoroid impacts, extreme temperature fluctuations, galactic cosmic rays, destructive dust, and seismic events. Loss of air or atmospheric leakage from a habitat poses safety challenges that demand proper attention. Such leakage may arise from micro-meteoroid impacts, crack growth, bolt/rivet loosening, and seal deterioration. In this paper, leakage estimation in deep space habitats is posed as an inverse problem. A forward pressure-based dynamical model is formulated for atmospheric leakage. Experiments are performed on a small-scaled pressure chamber where different leakage scenarios are emulated and corresponding pressure values are measured. An exponentially-weighted adaptively-refined search (EWARS) algorithm is developed and validated for the inverse problem of real-time leakage estimation. It is demonstrated that the proposed methodology can achieve real-time estimation and tracking of constant and variable leaks with accuracy.
\end{abstract}

\begin{keyword}
Deep space habitats \sep Leakage estimation \sep Real-time inversion \sep Exponential weighting \sep Adaptively-refined search 
\end{keyword}
    
\end{frontmatter}


\section{Introduction} \label{sec:sample1}
The quest to send humans to outer space and beyond has been a challenge for the human race for over 50 years. With technological advancements and renewed excitement in public at large, space-related research and development activities are growing at an aggressive pace. In 2015, NASA revealed long-term extraterrestrial settlement plans, which require the development of safe and functional space habitats \cite{maghareh2021role}. Such deep space habitats face unprecedented challenges of extreme environments like lack of breathable atmosphere, abnormal gravitational field, extreme temperature fluctuations, meteoroid impacts, intense radiation, seismic events, destructive dust, and so on \cite{landis1996dust,lammlein1977lunar,brown2015effects,crusan2018deep}. Such habitats must be resilient in that they can autonomously sense, anticipate, respond to, and learn from disruptions, and recover to a safe state in the minimum amount of time possible \cite{dyke2021strategies}. The loss of a breathable environment or air leakage threatens the safety of the crew and demands research into detection and assessment methods. 

Leakage identification or estimation is an emphasis of research in the oil/gas-related industries \cite{murvay2012survey}. The leakage characteristics of habitats are distinct from pipelines. This limits the implementation of existing physical models, experimental setups, and procedures. As per the authors' knowledge, there is no literature available on leakage estimation for deep space habitats. However, the research conducted around the international space station (ISS) can serve the purpose. Multiple sources are cited for air leakage from a space habitat, including micro-meteorite and particle impacts, crack growth, joint loosening, and seal deterioration \cite{wilson2008leak}. Atmospheric leakages in the ISS can arise locally (impacts and crack growth) or globally (seal deterioration). Different techniques have been studied for the ISS based on pressure, acoustic emission, and thermal imaging \cite{lemon1990technology}. Among different sensors on the ISS, pressure-based sensing instruments are considered sufficient for leakage detection \cite{wilson2008leak}. A pressure-based seal integrity monitoring system has been developed and evaluated for global atmospheric leakages on the ISS \cite{dunlap2007full}. However, to date, we have found no existing studies on local leakage estimation using pressure-based techniques.

Leakage identification (LI) can be posed as an inverse problem where the goal is to detect and quantify the cause of leakage from the effect \cite{rautela2021ultrasonic, rautela2021inverse}. In pressure-based LI, the cause of leakage is local damage in the form of a hole, and the effect is the pressure drop inside the habitat. LI can be categorized into three different levels, i.e., detection (Level-1), severity assessment (Level-2), and localization (Level-3) \cite{zaman2020review}. Pressure-based LI technique is global in nature, and it is utilized for leak detection and severity assessment. The technique has limited capabilities for localization in its raw form \cite{lemon1990technology}. 

In this work, we have revisited the gas dynamics model based on the depressurization of a storage reservoir of finite size \cite{keith1970calculation}. This reservoir is utilized for exploring the detection of leaks in the habitat. The forward model creates a functional relationship between the area of leakage and the pressure drop in the habitat. A pressure chamber/pressure box is commissioned with onboard instrumentation and equipment to study leakages. A flow controller is used to emulate leakages (constant and variable) by controlling the mass flow (proportional to the area of leakage). The flow controller provides a physical manifestation of an actual hole in the space habitat due to particle impact, crack growth, rivet/bolt loosening, and so on. A pressure transducer is also installed to measure the pressure inside the chamber due to the induced leakage. The pressure measurements are used for the forward model calibration to estimate and track leaks in real-time before they become threatening.

An exponentially-weighted adaptively refined search (EWARS) algorithm is developed and validated for the rapid and less noisy estimation of leakage area in real time. The algorithm is a combination of adaptively-refined search at every time step and exponential weighting of the objective functions over the time steps. The adaptively-refined search (ARS) is an improved brute-force or exhaustive search to find the optimal area of leakage while minimizing the mean square-based objective function. In order to reduce the number of computations as compared to the full brute-force search (fBFS), an adaptive refining scheme is proposed for every time step. However, the real-time leakage estimate becomes noisy if the exhaustive search is performed at every time step independent of other time steps. For less noisy estimates, the objective function at the current time step is added to previous functions that occurred at other time steps with exponentially decaying weights. 

The novelty and contribution of the paper can be broadly described as (a) commissioning of a pressure chamber with onboard instrumentation and equipment for leakage emulation and measurements, (b) reformulating the chamber depressurization model for leakage studies in habitats, (c) real-time model calibration-based inversion using EWARS scheme. The paper is organized as follows: Sec.~\ref{sec:theory} contains the theoretical foundation of the forward model and inversion schemes, Sec.~\ref{sec:experiments} discusses the experimental setup and details, Sec.~\ref{sec:results} contains results, Sec.~\ref{sec:discuss} has discussions, and the paper is concluded in Sec.~\ref{sec:conclusion}.

\section{Theoretical Background} \label{sec:theory}
\subsection{Forward physics-based model}\label{ssec:forward}
\begin{figure}
\centering
\includegraphics[width=0.6\textwidth]{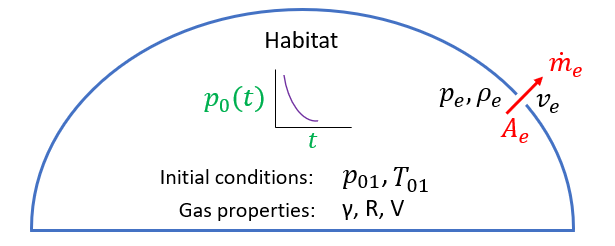}
\caption{Schematic representation of gas expulsion from a closed reservoir.}
\label{fig:forward}
\end{figure}

Depressurization of a tank is a well-studied problem in the literature. The gas dynamics-based model governing depressurization can also be used to model the leakage phenomenon in habitats. The model captures the physics of atmospheric air expulsion from a pressurized chamber through an orifice. Fig.~\ref{fig:forward} provides a schematic representation of gas expulsion from a closed container. 
This depressurization process produces a pressure change inside the reservoir. The pressure change can either be modeled through an unsteady state analysis or spatial lumped analysis \cite{owczarek1964fundamentals}. The difference is that in the latter model, the pressure change inside the reservoir is uniform and assumes this variation is independent of spatial coordinates \cite{john2006gas}. This assumption works well with associated applications \cite{nabi2000dynamic,winters2012experimental}. Therefore, we have utilized spatial lumped analysis for formulating the forward model in our study, which mathematically represents pressure, $P(x,y,z,t)$ as $P(t)$. Along with this, two more assumptions are incorporated into the mathematical formulations of Ref.~\cite{keith1970calculation,reynolds1958blowdown}. We have assumed no heat transfer in or out of the pressure reservoir (isentropic flow conditions) and ideal gas behavior. The equation of conservation of mass provides the rate of change of mass inside the reservoir and can be written as
\begin{equation} \label{eq:masscons}
\frac{dm}{dt} = -\dot{m_e} = \rho_e A_e v_e
\end{equation}
where $\dot{m_e}$, $\rho_e$, $A_e$, $v_e$ are mass flow, density of gas, area and velocity at the exit, respectively.

The initial mass of the gas inside the reservoir is $m=\rho_0V$, where $\rho_0$ and V are the density and volume of the gas inside the reservoir, respectively. The rate of change of the mass inside the reservoir can be written using isentropic relations between initial ($p_{01}$,$\rho_{01}$), reservoir ($p_0$,$\rho_0$), and exit ($p_e$,$\rho_e$) points ($p_0/\rho_0^\gamma$ = $p_{01}/\rho_{01}^\gamma$ = $p_e/\rho_e^\gamma$). Here, $\gamma$ is the ratio of specific heats.

\begin{equation}\label{eq:isentropic}
\frac{dm}{dt} = V\frac{d\rho_0}{dt} = \bigg(\frac{\rho_{01}V}{\gamma p_{01}^{1/\gamma}}\bigg) p_0^{\frac{1-\gamma}{\gamma}} \frac{dp_0}{dt}
\end{equation}

The velocity of the gas at the exit can be obtained using steady and adiabatic nozzle flow enthalpy conditions between reservoir and exit points $h_0 = h_e + 0.5v_e^2$. Also, invoking the ideal gas relationship, the velocity at the $v_e$ can be written in terms of reservoir temperature ($T_0$) and exit temperature ($T_e$) as
\begin{equation}\label{eq:enthalpy}
v_e^2 = 2c_p(T_0-T_e) = 2\frac{\gamma}{\gamma-1}\Bigg(\frac{p_0}{\rho_0}-\frac{p_e}{\rho_e}\Bigg)
\end{equation}

The variables present in the equation for exit mass flow $\dot{m}_e$ (Eq.~\ref{eq:masscons}) depend upon the pressure change in the reservoir. Depending upon the exit Mach number ($M_e$), the exit flow can be divided into two stages. The equations for Stage-I with choked flow ($M_e$=1 and $p_e>p_{atm}$) and Stage-II with unchoked flow ($M_e<1$ and $p_e = p_{atm}$) can be written as\\
\begin{align} \label{eq:stage}
\begin{aligned}
    \text{Stage-I}: p_0 &= \Bigg(\frac{\gamma+1}{2}\Bigg)^{\frac{\gamma}{\gamma-1}} p_e \\
    \text{Stage-II}: p_{atm} &\leq p_0 \leq \Bigg(\frac{\gamma+1}{2}\Bigg)^{\frac{\gamma}{\gamma-1}} p_{atm} 
\end{aligned}
\end{align}

Using Eqs.~\ref{eq:masscons}, \ref{eq:enthalpy}, \ref{eq:stage}, and the isentropic relations, the exit flow $\dot{m}_e$ during each stage is 
\begin{equation} \label{eq:mdot}
\dot{m_e} = \left\{
\begin{array}{ll}
    \frac{\rho_{01}}{p_{01}^{1/\gamma}}
    (\frac{2}{\gamma+1})^{\frac{1}{\gamma-1}}
    A_e
    \sqrt{\frac{2\gamma p_{01}^{1/\gamma}}{(\gamma+1)\rho_{01}}}
    p_0^{\frac{\gamma+1}{2\gamma}}
    & ,\text{Stage-I} \\
    \rho_{01} \big(\frac{p_{atm}}{p_{01}}\big)^{1/\gamma} 
    A_e 
    \sqrt{\frac{2\gamma}{\gamma-1}\frac{p_0}{\rho_{01}(p_0/p_{01})^{1/\gamma}} \big(1-(\frac{p_{atm}}{p_0})^{\frac{\gamma-1}{\gamma}}\big)} 
    & ,\text{Stage-II} \\
\end{array} 
\right.
\end{equation}

Eqs.~\ref{eq:masscons}, \ref{eq:isentropic} and \ref{eq:mdot} can rearranged to write the governing differential equations for each the two stages.
\begin{equation}\label{eq:gde}
\begin{gathered} 
\text{Stage-I}: 
\frac{dp_0}{dt} = C_I p_0^{\frac{3\gamma-1}{2\gamma}}, \text{where} \hspace{2mm} 
C_I = -\frac{\gamma A_e}{V}\bigg(\frac{2}{\gamma+1}\bigg)^{\frac{1}{\gamma-1}}
\sqrt{\frac{2\gamma p_{01}^{1/\gamma}}{(\gamma+1)\rho_{01}}} \\ 
\text{Stage-II}: 
\frac{d\bar{p}_0}{dt} = C_{II} \bar{p}_0^{\frac{\gamma-1}{\gamma}}
\sqrt{\bar{p}_0^{\frac{\gamma-1}{\gamma}}-1}, \text{where} \hspace{2mm} 
C_{II} = -\frac{\gamma A_e}{V}\bigg(\frac{2}{\gamma-1}\bigg)^{\frac{1}{2}}
a_{01}\bigg(\frac{p_{atm}}{p_{01}}\bigg)^{\frac{\gamma-1}{2\gamma}}
\end{gathered}
\end{equation}
where, $\bar{p}_0 = p_0/p_{atm}$, $a_{01} = \sqrt{\gamma R T_{01}}$ is the speed of sound at initial conditions, $R$ is the gas constant, and $T_{01}$ is the initial temperature. A more detailed mathematical formulation of the model is described in Ref.~\cite{john2006gas}.

Note that Eq.~\ref{eq:gde} gives a relationship for the rate of change of pressure $p_0$ that is a function of the area of the leakage $A_e$. The ordinary differential equation is first-order and non-linear. The stage-I equation has an analytical solution for a particular value of $\gamma$. The value of $\gamma$ is 1.4 for air. A fourth-order Runge-Kutta (RK4) technique can be used to calculate the solution for the stage-II equation numerically \cite{tan2012general}. Since the area of leak $A_e$ is unknown, the solution of either equation cannot be computed. Therefore, a model-calibration-based inversion scheme is required.

\subsection{Inversion scheme}
In the forward model (Eq.~\ref{eq:gde}), $A_e$ acts as an input and $p_0$ as the output. However, only $p_0(t)$ can be measured experimentally. So, the goal of the inversion is to estimate $A_e$ from $p_0(t)$ for the given physics-based forward model using the experimental measurements. Model calibration falls under inverse problems where unknown parameters (cause) are estimated using experimental observations of known parameters (effect) \cite{gupta2006model}.

In this study, we have followed an optimization framework and used a mean-squared error (MSE) based objective function, given as
\begin{equation}
    \arg \min_{A_{min}} F_t = [P_{phy}(A_e,\theta,t)-P_{exp}(t)]^2
\end{equation}
where, $P_{phy}(A_e,\theta,t)$ is the pressure calculated using the physics-based model (Eq.\ref{eq:gde}), which is a function of the area of leakage $A_e$, other constants $\theta$ and time $t$. Quantity $P_{exp}(t)$ are the real-time pressure measurements from the experiment. The goal is to find the value of $A_e = A_{min}$ that minimizes the MSE between the model and experimental data.

Optimization problems can be solved with either global or local approaches. Local techniques are based on gradient computations to move towards local optima \cite{ruder2016overview}. However, it is not possible to find gradients of the solution of Eq.~\ref{eq:gde}(II), because the solution itself must be calculated numerically at every time step when an appropriate $A_e$ is entered. Several global techniques which enable gradient-less search are available in the literature \cite{floudas2013state}. Since the focus of this inverse problem is on a single-dimensional search, a more straightforward brute-force or exhaustive search scheme is adopted. Brute-force search is based on the evaluation of the objective function over the search space \cite{tellez2019optimal}. The accuracy of the solution depends upon the discretization of the search space. Finer discretization yields better results on account of the larger number of computations and, consequently, more computational time. For rapid estimation of leakage area, we propose an adaptively-refined part of EWARS that will provide similar accuracy but with a lower number of computations than the fBFS. 
\begin{figure}
    \centering
    \includegraphics[width=0.9\textwidth]{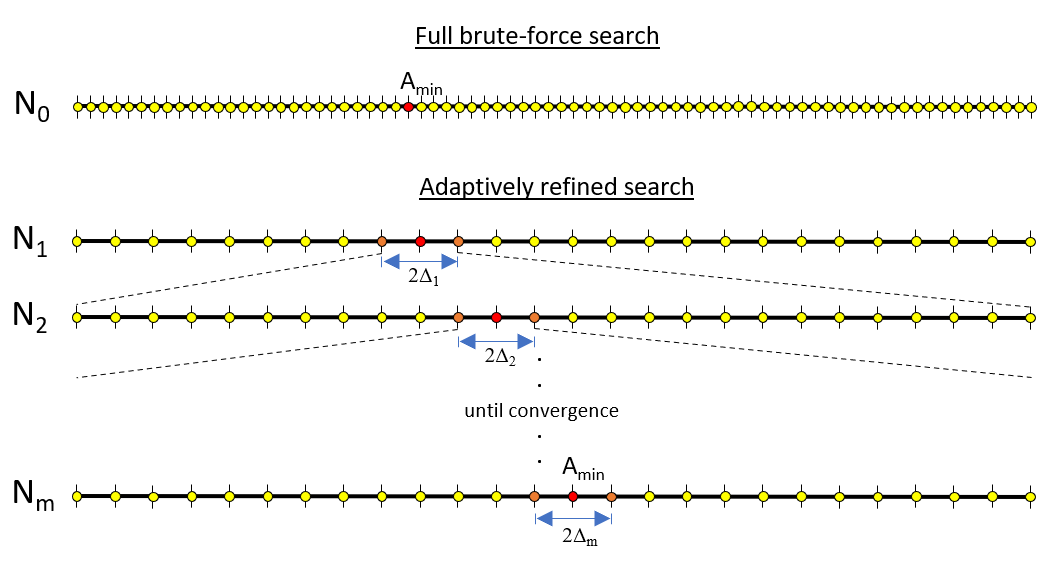}
    \caption{Pictorial representation of full brute-force search and adaptively refined search in one dimensional search space.}
    \label{fig:ARS}
\end{figure}

Fig.~\ref{fig:ARS} shows a pictorial representation of an adaptively-refined search. In the figure, $N_0$ represents the discretization of the full search space. In the fBFS search scheme (Fig.~\ref{fig:ARS}: topmost illustration), the MSE-based objective function is evaluated at all grid points, and the minima ($A_{min}$) is the argument of the function's minimum. However, in ARS, the search space is coarsely discretized at first, and the objective function is evaluated at all points. Then, the region around the minima (defined by $2\Delta$) is further discretized into finer grid points, and the minima is recalculated until a predefined convergence (on accuracy) is achieved. Quantities $N_1$, $N_2$,...$N_m$ in the figure represent discretizations of the search space at every step of ARS. For instance, fBFS demands $N_0$ = 10,000 (equivalent to 10000 computations) to achieve a particular $A_{min}$. By choosing proper $\Delta_1$ and $\Delta_2$ values, similar accuracy can be obtained with $N_1$ = 100 followed by $N_2$ = 200 (equivalent to 300 computations in total). The proposed ARS component of EWARS makes the search faster without compromising the accuracy of the inversion. One important point to note here is that the ARS guarantees convergence only when the objective function is convex in nature with one global optimum. For the problem at hand, the pressure gradient $p_0$ linearly depends on the area of leakage $A_e$ (refer to Eq.~\ref{eq:gde}). This means if the area of leakage increases, the pressure inside the reservoir drops. The MSE-based objective function introduces convexity.

The second component of the EWARS is exponential weighting (EW). Ideally, leakage estimation would be performed at every time step using ARS. However, experimental data has noise, which will affect the results if an independent ARS were performed at every time step. To address this practical concern, we have introduced exponentially decaying weights to the recurring objective function. Mathematically, the exponential weighting of objective function $F_t$, is 
\begin{gather}
    S_0 = F_0, \hspace{2mm} t = 0 \nonumber\\
    S_t = \alpha F_t + (1-\alpha) S_{t-1}, \hspace{2mm} t>0, \hspace{2mm} 0\leq\alpha\leq1
\end{gather}

In the above equation, $\alpha$ works as a smoothing factor, which ranges between 0 and 1. A higher value of $\alpha$ means there is a greater contribution of the current time step than previous time steps in the overall function $S_t$. The value of $\alpha$ defines the extent of exponential decay of previous time steps, and $\alpha$ can also be viewed as a memorization/forgetting factor. A higher value means a faster forgetting (or lower memorization) of the objective function at previous time steps. For instance, if $\alpha=1$, then $S_t = F_t$, which refers to maximal forgetting and no contribution of previous objective functions $S_{t-1}$ to the overall function $S_t$. In this case, any noise present in the pressure measurements propagates directly to the leakage area estimation without any smoothing or filtering. A smaller value of $\alpha$ seems desirable but will introduce a statistical bias. The desired value of $\alpha$ depends on the requirements of a given application and is thus problem-dependent \cite{kim2022ewma}. For shorter time-series data \cite{aparisi2004optimization}, a value of 0.2 is used and for a long time-series \cite{kim2022ewma}, $\alpha<0.001$ is recommended.

The complete EWARS algorithm is presented as pseudocode in Alg.~\ref{alg:algo}. The algorithm needs smoothing factor $\alpha$, convergence criteria $\epsilon$, number of grid points $N$, constants of Eq.~\ref{eq:gde}, initialization of objective function along with upper and lower bounds on the area. The algorithm is able to run online using pressure measurement data from the experiments. At every time step, the ARS convergence criteria need to be fulfilled.

\begin{algorithm}[h!]
\caption{EWARS algorithm}
\begin{algorithmic}[1]
    \State Data: $\alpha$, $\epsilon$, $N$, $\theta$
    \State Initialization: $S_0 = 0, [A_{lb},A_{ub}]$
    \State Real-time measurement: $P_{exp}(t)$
    \For{t = start$\rightarrow$end}
    \Repeat                                                     \Comment{Begin ARS}
    \State $A_e \gets A_{lb}$:$N $:$A_{ub}$
    \State $\Delta = (A_{ub}-A_{lb})/N$ 
    \State $P_{phy}(A_e,\theta,t) \gets \dot{p}_0 = C p^{c}_0$  \Comment{Solve Eq.\ref{eq:gde}}
    \State $F_t \gets [P_{phy}(A_e,C,t) - P_{exp}(t)]^2$        \Comment{Obj. function}
    \State $S_t \gets \alpha F_t + (1-\alpha) S_{t-1}$          \Comment{EW}           
    \State $A^* \gets \underset{A^*}{\arg \mathrm{min}}$  $S_t$ \Comment{Search}
    \State $[A_{lb},A_{ub}] \gets A^* \pm \Delta$       
    \Until{$\Delta\leq\epsilon$}                                \Comment{End ARS}
    \State $A_{min} \gets A^*$
    \EndFor    
\end{algorithmic}
\label{alg:algo}
\end{algorithm}

\section{Experimental setup}\label{sec:experiments}
\begin{figure}[h!]
\centering
\includegraphics[width=0.85\textwidth]{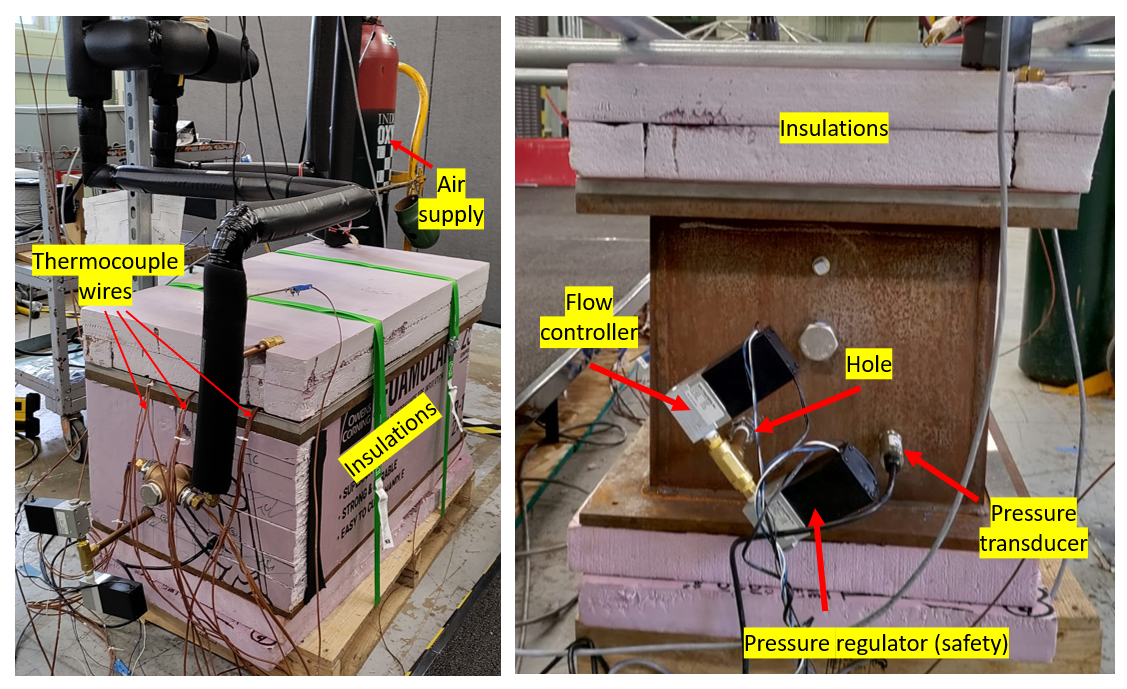}
\caption{Labeled snapshot of experimental setup (pressure chamber) and instrumentation.}
\label{fig:setup}
\end{figure}

A pressure chamber is commissioned for leakage estimation experiments. The purpose of the experimental setup is, in part, to study the leakage phenomenon and leak estimation from the point of view of future space habitats. The design of future habitats will be entirely different from the pressure chamber. However, this setup is sufficient to study the leakage phenomenon in isolation. The pressure chamber is constructed from six welded 1020 cold-rolled steel plates with a thickness of 12.5 cm (0.75 in). The approximate dimensions of the pressure box are 800 x 400 x 400 mm. Our facility cannot host vacuum-like ambient pressure. Therefore, we have conducted leakage experiments with 2 atm pressure inside the chamber while the outside pressure remains at 1 atm. The experimental setup is shown in Fig.~\ref{fig:setup}. 

The pressure chamber is covered with foam insulation to maintain the isentropic assumptions used in the forward model. Table~\ref{tab:intruments} shows the instrumentation and equipment installed in the pressure chamber to facilitate the experiments. A mass flow controller is used as a leakage emulator, which receives commands from a real-time target machine. The command lies between 0-10V, where 0V corresponds to a closed condition, and 10V represents the maximum mass flow. The pressure inside the chamber is measured with a pressure transducer. The temperature is measured via a thermocouple. The pressure regulator is used as a pressure safety check of the chamber. 

\begin{table}[h!]
\centering
\caption{On-board instrumentation and associated equipment}
\addtolength{\tabcolsep}{-1pt} 
\begin{tabular}{|l|c|c|} 
\hline
Instrument/Equipment & Manufacturer & Model \\ 
\hline
Mass flow controller & Kelly pneumatics & MFCL-A1012-010010-10LPM-4 \\ 
Pressure transducer & Omega & PX309-050A10V \\ 
Pressure regulator & Kelly pneumatics & LFR-15TK-05010-R4 \\ 
Thermocouple & IPS \& Omega & T-20-TT \& PFT2NPT-4T \\
Thermocouple probe & IPS & TG20T0142U00600MP \\
Target machine & Speedgoat & Performance real-time target machine \\
Data acquisition unit & m+p Vibrunner & - \\
\hline
\end{tabular}
\label{tab:intruments}
\end{table}

The instruments are calibrated and connected with the m+p Vibrunner data acquisition system and Speedgoat real-time machine for offline analysis and online estimation, respectively. Fig.~\ref{fig:ARS} shows the full architecture of the experimental setup along with connections. The offline analysis uses data collected with the m+p Vibrunner. This data is subject to anti-aliasing filters and thus is gathered at a higher sampling frequency (20 kHz) for noise characterization. The Speedgoat machine and Simulink I/O blockset interface enable online leakage estimation and use a lower sampling frequency of 1 kHz. 
\begin{figure}[h!]
    \centering
    \includegraphics[width=1.0\textwidth]{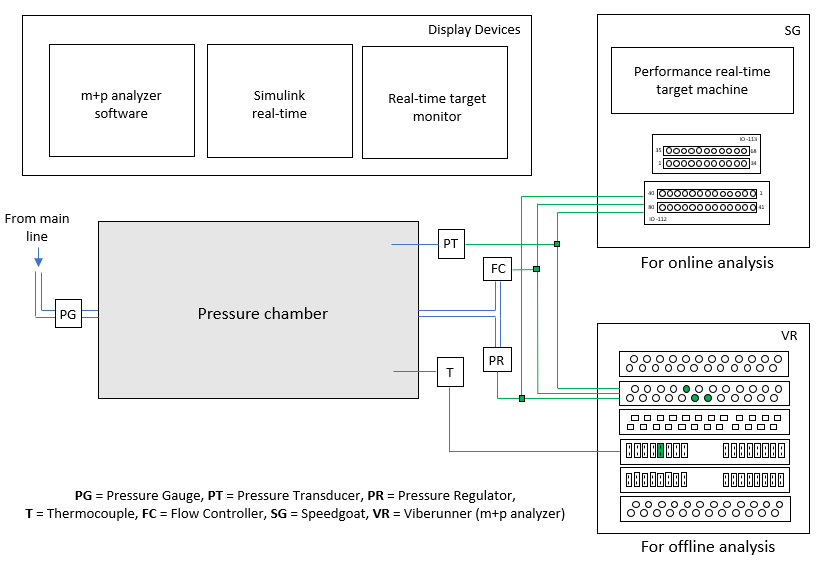}
    \caption{Schematic diagram of the experimental setup, instrumentation and equipment.}
    \label{fig:connect}
\end{figure}

\section{Results} \label{sec:results}
There are two different viewpoints on the parametric representation and associated units of a leak. The oil and gas industry adopts $\dot{m}$ ($\text{kg/m}^3$) as the representative parameter of a leak because of the nature of the fluid flow through a channel \cite{murvay2012survey}. However, in our application, fluid behavior and control volume are different. The cause of leaks in a space structure could be meteorite impacts, crack growth, rivet/bolt loosening, and so on. Therefore, a geometrical representation of a leak is more practical, i.e., the area of the leak. In our experimental setup, a more advanced mass flow controller emulates leakage by using a feedback controller to achieve a particular $\dot{m}$ corresponding to a command falling between 0-10 Volts. The calibration chart provided by the manufacturer converts leakage in Volts to leakage in standard liters per minute (slpm). This calibration is non-linear in nature, and we have fit a sixth-order polynomial to convert Volts into slpm. Then, Eq.~\ref{eq:mdot} can be used to transform $\dot{m}$ (slpm) to $A_e$ (square metres or square millimetres). It can be seen that the $\dot{m}$ depends on the area of the leak ($A_e$) and pressure ($p_0$). It is observed that the flow controller starts increasing the area for larger leaks to maintain a constant mass flow. For leaks below $\dot{m}$ of 6V ($A_e$=0.28 $\text{mm}^2$), the change in the area is less than 5 percent for a period of 3 minutes and hence can be considered constant. Therefore, we have utilized the flow controller in the 0-6V range with the corresponding area of leak between 0-0.28 $\text{mm}^2$.

The atmospheric pressure ($p_{atm}$) of air are standard sea-level values. The value of initial temperature ($T_{01}$) and pressure ($p_{01}$) is measured at approximately around 300K and 2 atm, respectively. The density ($\rho_{01}$) can be calculated using the ideal gas law. The search space is constrained with upper and lower bounds on the leak area as [1e-3,1] $\text{mm}^2$ instead of an unconstrained search in [0,$\infty$]. The value of $\epsilon$ for EWARS is chosen as 5e-5 $\text{mm}^2$.

\subsection{Constant Leak}
Each test begins with a zero command to the mass flow controller. Then, a step function is applied at a predetermined time from the Simulink interface to the flow controller to emulate a constant leak. The leakage causes the pressure inside the chamber to drop below 2 atm. Pressure and initial temperature measurements inside the pressure chamber are collected in real time with the Simulink interface. Pressure measurements outside of the chamber are assumed to remain at 1 atm. The data enters the EWARS module, which estimates the leakage area (in $\text{mm}^2$) online and provides a visualization. Value of $N$=150 and $\alpha$=0.125 are used for a constant leak. 

\begin{figure}[h!]
    \centering
    \includegraphics[width=0.65\textwidth]{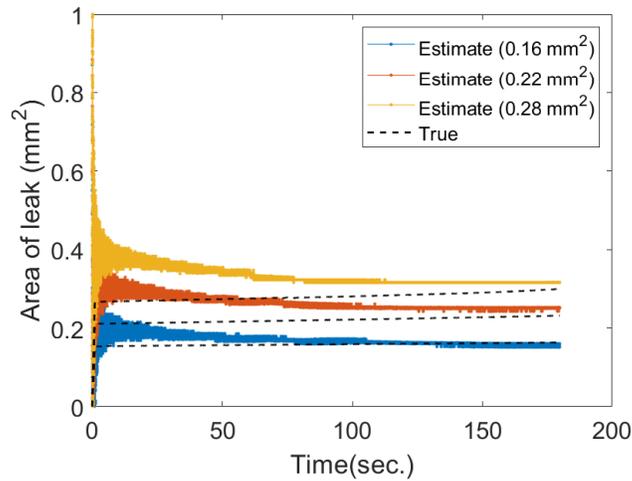}
    \caption{Constant leakage estimation with EWARS}
    \label{fig:constant}
\end{figure}

Figure~\ref{fig:constant} shows the true value and the estimated area for three different cases with constant leakage values (leak area of 0.16, 0.22, and 0.28 $\text{mm}^2$). It is observed from the figure that the algorithm is able to converge toward the true value. The difference between the true and estimated value improves with time. It is seen that the EWARS-based leak estimation takes less time for lower leaks and more time for higher leaks. It is because higher leaks correspond to a larger pressure drop. It is also observed that as time passes, the separation between different intensities of leakages becomes evident. An approximate time of 2.5-3 minutes is required for the separation and convergence of the estimated leakage to a true value. This time requirement applies to other leakage intensities as well. 

\subsection{Variable leak}
For variable leaks of both gradually increasing or decreasing nature, multi-step functions are used as the inputs to the flow controller. The goal of conducting variable leak scenarios is to study the hole elongation (increasing leak) and hole patching (decreasing leak) scenarios. There can be multiple reasons for an increasing leak in deep space habitats. Some of them are differential temperature gradients and fatigue loading. However, a decreasing leak is possible through an intervention of an agent (human or robot) to fix the leakage. The repair dynamics of the agent will define the dynamics of pressure coming back to the normal state. In order to emulate increasing and decreasing leaks, we have used multi-step functions imitating the slow dynamics of variable leaks.

\begin{figure}[h!]
\begin{minipage}[b]{1.0\textwidth}
    \centering
    \includegraphics[width=0.75\textwidth]{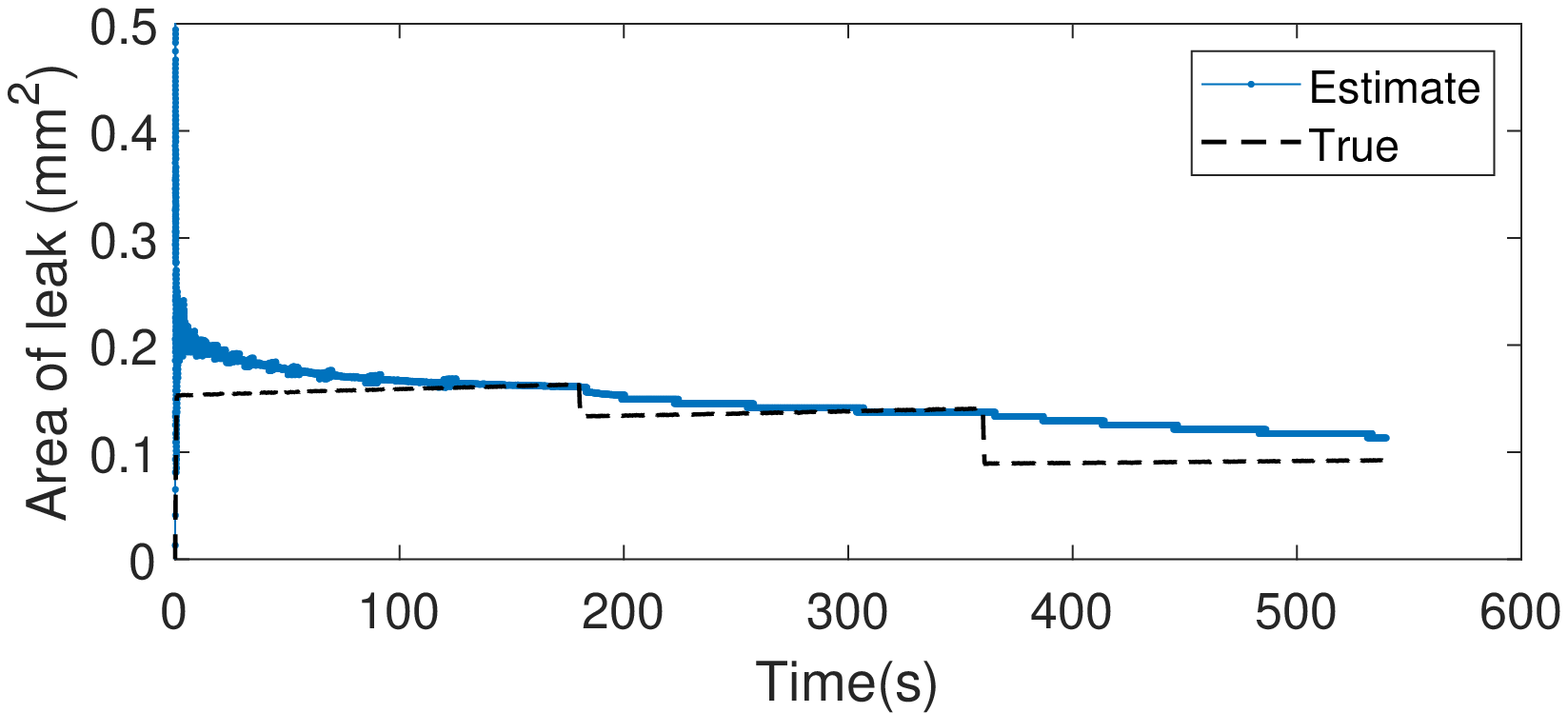}
\end{minipage}
\begin{minipage}[b]{1.0\textwidth}
    \centering
    \includegraphics[width=0.75\textwidth]{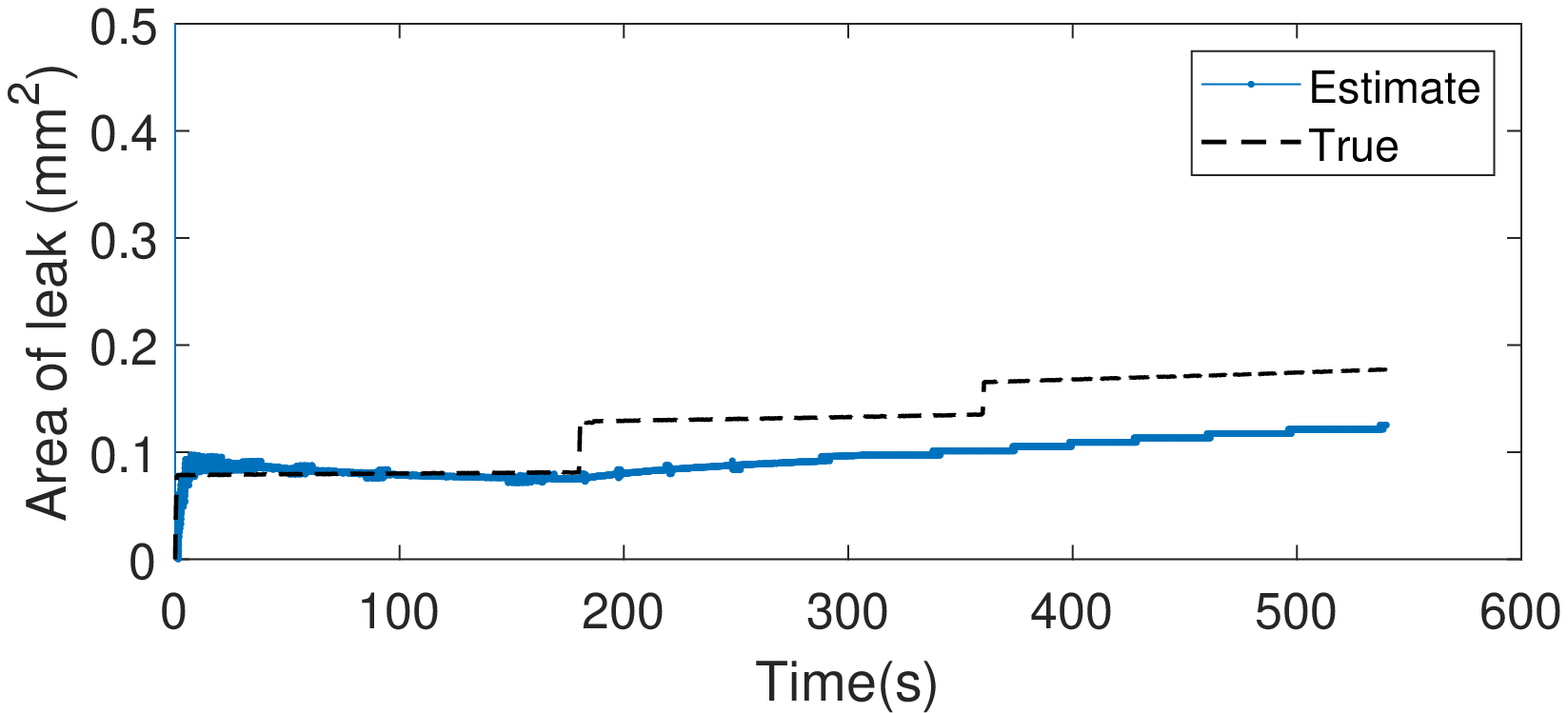}
\end{minipage}
\caption{Variable leakage estimation using EWARS, (top): Decreasing leak, (bottom): Increasing leak}
\label{fig:varleak}
\end{figure}

For gradually increasing leaks, the leak magnitude is varied from 0.08 to 0.16 $\text{mm}^2$ in 3 steps with a time interval of 3 minutes for every step. Similarly, for gradually decreasing leaks, the magnitude is reduced from 0.16 to 0.08 $\text{mm}^2$ in 3 steps. The real-time pressure measurements are fed into the EWARS algorithm. The value of $N$ is selected as 250 along with $\alpha$ = 0.01. The value of $\epsilon$ and initial bounds on the area is the same as the constant leak case. The results are presented in Fig.~\ref{fig:varleak}. It can be seen that the EWARS algorithm can estimate and track a variable leak. The true and estimated value closely matches.

From the figure, it can be observed that the estimation accuracy is better for a decreasing leak. This improvement is because the pressure drop slows down for a decreasing leak and becomes easier for EWARS to track it in real-time. On the other hand, the pressure drop surges in the case of an increasing leak, which gives EWARS a hard time tracking it. Similar behavior is seen when EWARS was able to estimate lower leaks faster than the higher leaks.

\section{Discussion} \label{sec:discuss}
In the previous section, we experimentally validated the capability of the EWARS scheme for estimating the size of leaks in a pressure reservoir emulating a space habitat. The algorithm is defined by parameters $\epsilon$, $N$, and $\alpha$. For a constant $\epsilon$ for our problem, the scheme can be written and represented with parameters as EWARS($N$,$\alpha$). The EWARS scheme is faster (defined by $N$) and provides less noisy estimates (defined by $\alpha$) as compared to fBFS. Table.~\ref{tab:comparetime} compares the computational time of fBFS and EWARS when run in an offline mode. Both algorithms are run on a personal computer with an intel-core-i9-10920X CPU and 32 GB of RAM.

\begin{table}[h!]
\centering
\caption{Comparison of the computational time (minutes) for fBFS and EWARS}
\addtolength{\tabcolsep}{-1pt} 
\begin{tabular}{|l|c|c|} 
 \hline
 Scheme/Scenario & Constant leak & Variable leak \\ 
 \hline
 fBFS & 8.67 & 31.60  \\ 
 EWARS & 1.27 & 4.79 \\
 \hline
\end{tabular}
\label{tab:comparetime}
\end{table}

\begin{figure}[h!]
    \centering
    \includegraphics[width=0.65\textwidth]{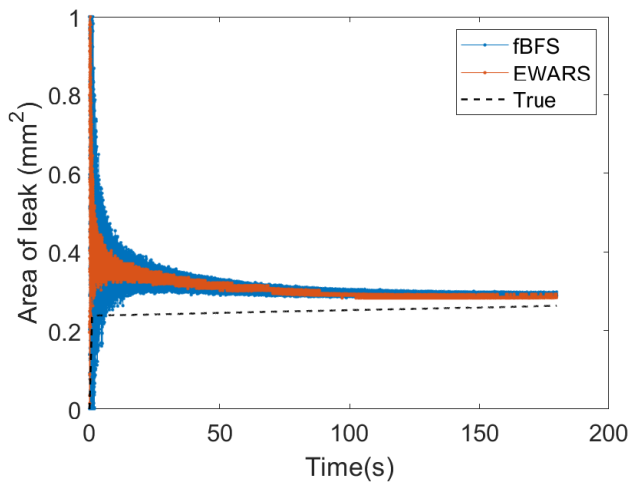}
    \caption{Comparison of the estimates from fBFS and EWARS(150,0.125)}
    \label{fig:compareleak5V}
\end{figure}

Fig.~\ref{fig:compareleak5V} shows fBFS and EWARS estimates for a 0.25 $\text{mm}^2$ leak with $\alpha$=0.125. It is seen from the figure, EWARS provides a less noisy estimate than fBFS.

EWARS scheme is a model-calibration technique that minimizes the MSE between the forward physics-based model and experimental data. During the implementation of EWARS for all leak intensities and scenarios, the mean absolute percent error is less than 1\%. These results directly reflect the model's performance and inversion technique developed herein. We have also seen that the error between the experimental data and the model increases for larger leaks. For instance, with a 0.08 $\text{mm}^2$ leak, the mean absolute percent error is less than 0.5\%, whereas for a 0.28 $\text{mm}^2$ leak it is 1\%. Possible reasons for such residual errors are model discrepancies and uncertainties associated with related parameters and initial conditions. The physics-based model follows a spatial lumped approach and is built under the assumptions of isentropic flow and ideal gas conditions. Also, the model does not account for different types of losses. One of them could be friction during the expulsion of air at the exit. The atmospheric pressure ($p_{atm}$) and temperature ($T_{atm}$) conditions are aleatoric in nature. Manual chamber filling operation introduces another source of aleatoric uncertainty in the initial conditions ($p_{01}$,$T_{01}$).

\section{Conclusion} \label{sec:conclusion}
In this paper, we have investigated the pressure-leakage phenomenon from the point of safety of deep space habitat. We have commissioned a pressure chamber in the laboratory to perform isolated leakage experiments. The pressure chamber is instrumented with sensors, a data acquisition system, a real-time target machine, and a real-time Simulink control interface. A novel online inversion scheme EWARS is developed and experimentally validated for rapid (ARS part) and less noisy (EW part) estimation of the leakage area from pressure measurements. The results show that EWARS was able to track and estimate constant leaks of different intensities as well as variable leaks (increasing and decreasing leaks). The implementation of EWARS to online inverse problems can be controlled via $\epsilon$, $\alpha$, and $N$ provided the objective function is convex in nature. The experiments were conducted under isentropic conditions ($\Delta Q$ = 0 and no losses). As a part of future work, another more realistic habitat system is being commissioned where leakage estimation will be studied with heat transfer scenarios. The uncertainty analysis and quantification associated with the physics-based model and experiments will be considered in future investigations.

\section{Acknowledgements}
This work was supported in part by a Space Technology Research Institutes grant (number 80NSSC19K1076) from NASA’s Space Technology Research Grants Program. The first author, Mahindra Rautela, acknowledges his fellowship grant under the Purdue-India research collaboration program through Science and Engineering Research Board (SERB)-India. 

\bibliographystyle{elsarticle-num} 
\bibliography{cas-refs}

\end{document}